# FAZSeg: A New User-Friendly Software for Quantification of the Foveal Avascular Zone

*V. K. Viekash[1], Janarthanam Jothi Balaji [2], Vasudevan Lakshminarayanan[3]*

[1]Department of Instrumentation and Control Engineering, National Institute of Technology, Tiruchirappalli, Tamil Nadu, 620015, India.
[2]Department of Optometry, Medical Research Foundation, Chennai, Tamil Nadu, 600 006, India
[3]Theoretical and Experimental Epistemology Lab, School of Optometry and Vision Science, University of Waterloo, Waterloo, Ontario N2L 3G1, Canada

## Abstract

### Introduction

Various ocular diseases and high myopia influence the anatomical reference point Foveal Avascular Zone (FAZ) dimensions. Therefore, it is important to segment and quantify the FAZs dimensions accurately. To the best of our knowledge, there is no automated tool or algorithms available to segment the FAZ's deep retinal layer. The paper describes a new open-access software with a user-friendly Graphical User Interface (GUI) and compares the results with the ground truth (manual segmentation).

### Methods

Ninety-three clinically healthy normal subjects included 30 emmetropia and 63 myopic subjects without any sight-threatening retinal conditions, were included in the study. The 6mm x 6mm using the Angioplex protocol (Cirrus 5000 Carl Zeiss Meditec Inc., Dublin, CA) was used, and all the images were aligned with the centre of the fovea. Each FAZ image corresponding to dimensions 420x420 pixels were used in this study. These FAZ image dimensions for the superficial and deep layers were quantified using the New Automated software Method (NAM). The NAM-based FAZ dimensions were validated with the manually segmented (ground truth) method.

### Results

The age distribution for all 93 subjects was 28.02 ± 10.79 (range, 10.0 – 66.0) years. For normal subjects mean ± SD age distribution was 32.13 ± 16.27 years. Similarly, the myopia age distribution was 26.06 ± 6.06 years. The NAM had an accuracy of 91.40 %. Moreover, the NAM on superficial layer FAZ gave a Dice Similarity Coefficient (DSC) score of 0.94 and Structural Similarity Index Metric (SSIM) of 0.97. On the other hand, the NAM on deep layers FAZ gave a DSE score of 0.96 and SSIM of 0.98.

**Conclusion**

A clinician-friendly GUI software was designed and tested on the FAZ images from deep and superficial layers. The new algorithm outperformed the device's inbuilt algorithm when measuring the superficial layer. This open-source software package is in the public domain and can be downloaded from https://github.com/VIEKASH2001/FAZSeg and can be used to measure the FAZ dimensions in the clinic.

**Introduction**

The Foveal Avascular Zone (FAZ) is located within the central macular region[1], where the retinal capillaries are absent.[2] This zone has the highest cone density[3,4] and overlays the inner retinal tissue.[4] This region is of clinical importance since the vascular arrangement around the fovea changes depending on the eye's (disease or refractive) status. The dimension of FAZ has been used as retinal markers in many ophthalmic conditions[1,5], and it also helps in prognosis.[6] For example, the FAZ size shows a relationship with the severity of Diabetic Retinopathy (DR), one of the major blinding diseases of the eye.[6] Multiple studies have reported enlargement of the FAZ area in subjects with diabetic retinopathy, and it is irregularly contoured with scattered microaneurysms.[7] High myopia also affects the size and shape of the FAZ.[8,9]. Therefore, it is important to segment and quantify the FAZs accurately.

Age, gender, refractive status, Axial Length (AXL), and central retinal thickness influence the FAZ dimensions.[10] For example, Tan et al. be consistent in referencing – give an upper number reported that the superficial layer and deep layer FAZ area showed a significant variation among the healthy normal young adults.[10] In myopic eyes with increased AXL, a reduced vascular density was reported both in the superficial and deep retina layers in the macular region.[11]

Several studies have reported FAZ data based on optical coherence tomography angiography (OCTA) images.[5,7,9,12] There are many ways in which the FAZ can be segmented in OCTA generated images.[13,14] Clinicians can use manual or system-generated markings. Manual marking is desired and is regarded as the ground truth. However, it is not only time consuming but also depends upon the availability of a trained clinician. Many use the automated inbuilt system-generated FAZ segmentation and the resultant dimensions. For example, the Cirrus 5000 Angioplex OCT device (Carl Zeiss Meditec Inc., Dublin, CA) and AngioVue (Optovue, Inc., Fremont, CA) segment the FAZ boundary.[1,12] A better segmentation method than that provided by the Cirrus 5000 Angioplex has been proposed by Agarwal et al.[15] The results of this study were compared with the ground truth and an extensive list of parameters defining the FAZ dimensions, and it was shown that it was possible to segment with 98.01 % accuracy.[15] Additionally, almost all the inbuilt automated methods provide FAZ dimensions of only the superficial layer FAZ and not the deep layer FAZ dimensions. To the best of our knowledge, there are no software tools available to segment the deep layer FAZ. This paper describes an easy to use, GUI based method and compare the results with the clinical method.

**Methodology**

**Subjects**

This retrospective study included subjects who had undergone comprehensive ophthalmic examinations along with an Optical Coherence Tomography (OCT) at a tertiary eye care centre in South India between January 1, 2019, and December 31, 2020. The study was conducted following the tenets of the Declaration of Helsinki and was approved by the Institutional Review Board of the Vision Research Foundation, Chennai, India. Exclusion criteria included any prior history or clinical evidence of retinal or systemic vascular disease, visual acuity worse than 6/9, any glaucoma, ocular hypertension, amblyopia and any ocular surgery. The total number of images included in the study was 93 (59 males, 34 females). The mean ± SD age was 28.02 ± 10.79 years. Myopic eyes were classified using a standardized severity scale based on mean spherical equivalent (MSE) in this study.[16] Group 1 consisted of individuals with emmetropia (> - 0.5D to +0.50 D SE), Group 2, low-moderate myopia (≤ -0.50 D to >-6.00 D SE), Group 3, high myopia (<-6.00 D SE).[16] The MSE was calculated from the manifest refraction during the ophthalmic evaluation. The best-corrected visual acuity (BCVA) was converted from Snellen's visual acuity. The intraocular pressure was measured using the Goldmann Applanation Tonometer (GAT) and acquired the standard clinical protocol with a calibrated Haag-Streit slit lamp.[17]

**Ocular biometry**

All participants' ocular biometric data were obtained using the non-contact and high-resolution optical biometric device IOLMaster 700 (Carl Zeiss Meditec AG, Jena, Germany). The IOLMaster 700 combines swept-source optical coherence tomography with a tunable laser wavelength centring on 1055 nm (a wavelength ranging between 1035 and 1095 nm) and a multi-dot keratometer.[18] This instrument performs optical cross-sectional B-scans to measure the ocular axial parameters and visualize axial anatomical structures as a two-dimensional OCT image. In addition to the overall Axial Length (AXL), the instrument provided the following parameters. 1. Corneal curvature (K-Reading), in both millimetre (mm) and dioptre (D), 2. Central Corneal Thickness (CCT) in microns, 3. Anterior Chamber Depth (ACD) in mm, 4. Lens Thickness (LT) in mm. 5. White To White (WTW) measurement direct measurement of the corneal diameter in mm.[18] The vitreous chamber depth is calculated by subtracting CCT, ACD, LT values from the AXL.[19,20] The Axial Length/Corneal radius Ratio (ACR) was calculated by the method given by Badmus et al. (2017).[21]

**Image specifications**

The FAZ was imaged using the commercially available Spectral-Domain OCT (Cirrus 5000, Carl Zeiss Meditec Inc., Dublin, CA). Angiography 6 x 6 mm program was used, and all the images were aligned with the centre of the fovea. All OCTA images were 8-bit grayscale images of 200 x 200 pixels corresponding to 6 mm x 6 mm (420 x 420 pixels). Each OCTA image was segmented by a clinician and used as the ground truth. This was compared with the new automated GUI method.

**Processing algorithms**

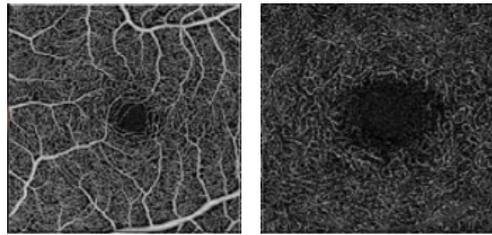

*Fig 1: FAZ images: Superficial layer(left) and deep layer (right)*

Clinically, the OCTA imaging is performed for two standard layers: The superficial and deep layers. Figure 1 shows the different layers of FAZ images. The primary goal of this research is to develop a software-based tool that can do the segmentation of superficial and deep layer images by processing non-marked images and then performing parametric calculations on the detected boundary. The proposed methodology was built based on a previously published algorithm[15]. The MATLAB R2020a App was used to implement the algorithm (MathWorks, Inc., Natick, MA).

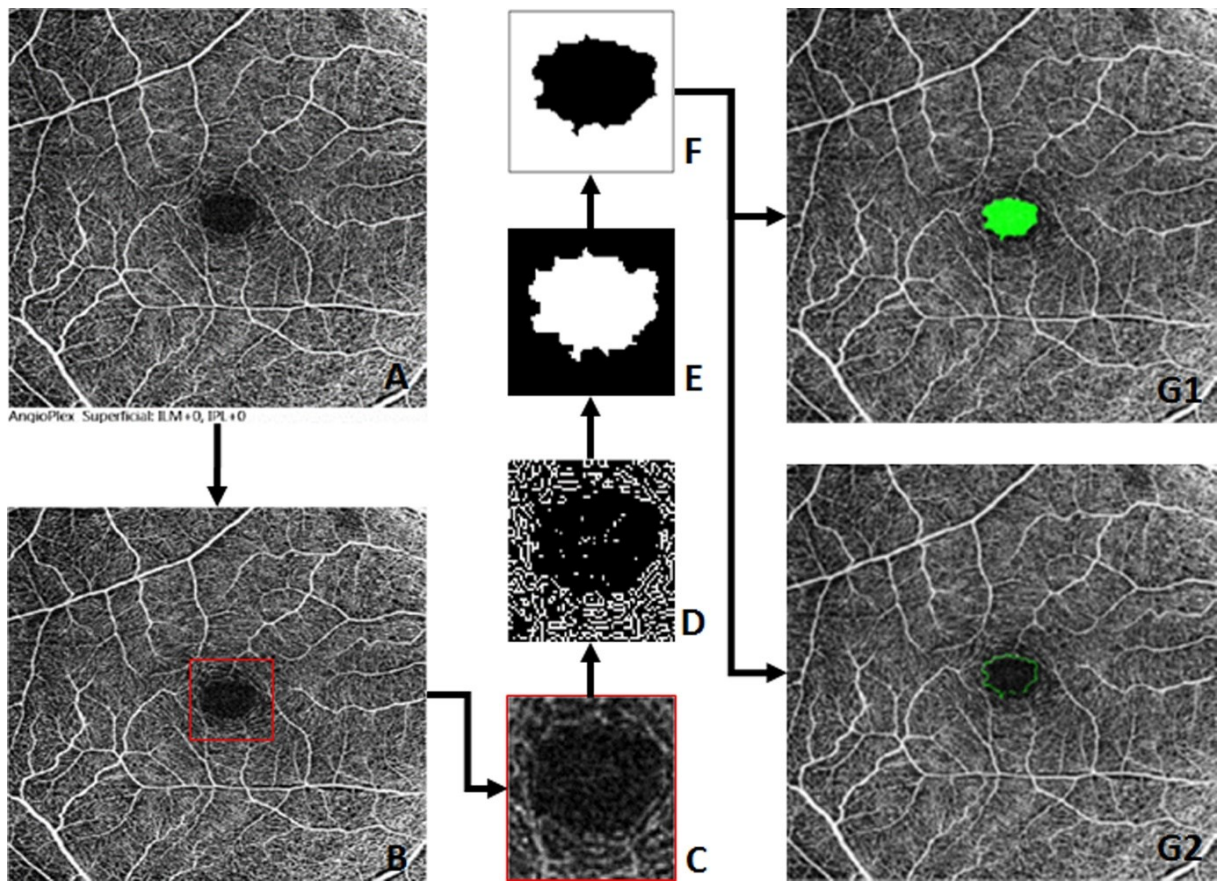

*Fig 2: Methodology flowchart: A: Original Non Marked Image, B: Cropped off the image description, C: Crop to the region of interest, D: Prewitts edge detection, E: Image dilation, F: Image erosion and false positive removal, G1: Infilled segmentation of detected FAZ region, G2: Outline segmentation of detected FAZ region.*

As illustrated in Figure 2, the non-marked image from which the FAZ region must be segmented is loaded first. The image description is then cropped off, and the FAZ region is plotted using the resulting image. The image is cropped around the region of interest to remove most of the vascular structures around the FAZ region for precise detection of FAZ boundaries and to avoid false-positive segmentations. Further, the Prewitt edge detector detects horizontal and vertical edges, thus distinguishing the region of interest's boundary from its surroundings. Unlike in Agarwal et al,[15] where two different threshold values were applied on prespecified image portions of a superficial layer image, a single thresholding parameter was used in our algorithm. The objective of unifying the parameter was to make it more generic and reduce the number of parameters that govern the threshold action. Also, the percentage of the area the centre portion covers varies with the depth of the layer. Hence a single thresholding parameter was used. In Agarwal et al,[15], the original image is complemented, and the Gaussian filter smoothes and minimizes noise.

In comparison to darker vessels, it also accentuates the bright FAZ region. However, the result of applying an edge detector to a pixel on edge is a vector that points across the edge from darker to brighter values, which remains constant as long as the gradient between pixels is maintained. As a result, it is clear that when complements and differences of Gaussian filtering procedures are applied to an image, only the intensity values of pixels change, not the differences/gradients between adjacent pixels. Thus these two steps are eliminated before applying the Prewitt edge detection operator. Once the image's edges have been recognized, the noise caused by nearby vasculature is reduced, which might impair segmentation by causing false-positive detections. A closure procedure involving erosion and dilatation can be used to remove the vascular structure. Also, this can aid in the creation of a FAZ zone that is accurately divided. In this study, a line-shaped morphological feature was used to dilate the image at angles of 0º, 45º, and 90º. The angles chosen were evenly spaced, which aids in the preservation of the FAZs' shape.

Furthermore, a disk-shaped element is used to accomplish image closure and False Positive Removal. The FAZ region is now clearly separated, but the other segmented regions around the FAZ must be removed due to the risk of false positives. Because all false positives were discovered to be significantly smaller than the FAZs, the largest of the detected boundary was chosen. FAZ zones are identified and segmented using two types of markings: infill and outline segmentation. Furthermore, parametric dimensional measurements were made.

The Agarwal et al[15] algorithm had 16 tunable variables that significantly affected the segmentation output. As a result, the user must provide 16 different values in order to complete accurate segmentation. However, having 16 variables can be a limitation when building a tool intended to assist segmentation quickly and with optimal flexibility. Thus, the variable sensitivity analysis tests were running to see the contribution of each of the 16 variables is to the final result. From this, it was concluded that 5 high-sensitivity tuneable variables are necessary to be used, and the rest 11 low-sensitivity tuneable variables were either eliminated or fixed at a constant value. This interface is illustrated in the GUI (Figure 3). Details of the interface is provided in the appendix.

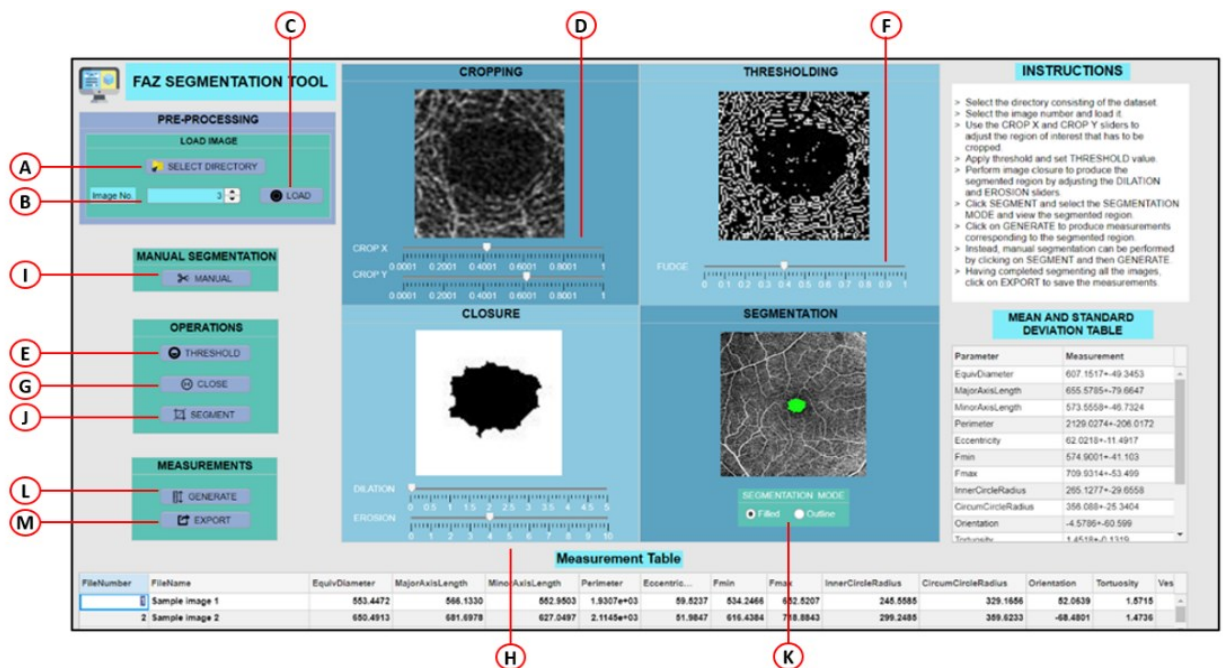

*Fig 3: Overall tool GUI design. A) Select dataset directory B) Image selection spinner C) Load image button D) Crop X & Crop Y sliders E) Threshold button F) Fudge factor slider G) Close operation button H) Dilation & Erosion sliders I) Manual segmentation J) Segment button K) Filled segmentation L) Outline segmentation M) Export measurements*

From this algorithm, 15 parametric dimensions of the FAZ can be measured.[22] These are: Area ($mm^2$), Diameter diameter (mm), Major axis length (mm), Minor axis length (mm), perimeter (mm), Eccentricity (mm), $F_{min}$ (mm), $F_{max}$ (mm), Inner circle radius (mm), Circumcircle radius (mm),

Orientation ($^0$) Tortuosity†, VAD†, VDI (mm), and Circularity index†. These parametric dimensions are compared between the Inbuilt Algorithm Method (IAM), Clinical expert method (CEM), and New automated method (NAM) images to analyze the algorithm's robustness.

**Statistical analysis**

All statistical analyses were performed using SPSS version 20 (SPSS Inc, Chicago, Illinois, USA). Variable normality was evaluated using the Kolmogorov-Smirnov test. Comparisons between the three different methods and the degree of myopia were performed using One-Way ANOVA and Kruskal Wallis statistical tests. The Pearson correlation analysis analyzed the correlation between the CEM vs NAM and IAM on all the FAZ dimensions. The Bland-Altman plots were used to compare agreement between CEM and NAM on the most commonly used FAZ dimensions, namely area, perimeter, and circularity index.

**Results**

Ninety-three eyes of healthy young adults' superficial and deep layer FAZ images were analyzed. This includes 30 eyes emmetropes, 31 eyes of low-moderate myopes, and 32 high myopes. The results are tabulated in Table 1. All results are expressed as the mean ± standard deviation and median (IQR) since some data were non-normally distributed. The demographic data and ocular biometry data were compared between emmetropic and myopic groups. Other than age and CCT, all other parameters were showed a statistically significant difference (Table 1)

*Table 1: Demographic data and ocular biometry details.*

|  |  | Age (years) | AXL (mm) | CCT (microns) | ACD (mm) | LT (mm) | VCD (mm) | CRC (D) | SE (D) | BCVA (LogMAR) | ACR† |
|---|---|---|---|---|---|---|---|---|---|---|---|
| Emmetropia (30) | Mean | 32.13 | 23.47 | 524.28 | 3.36 | 3.93 | 15.65 | 42.99 | 0.08 | -0.02 | 2.99 |
|  | SD | 16.27 | 0.86 | 27.03 | 0.37 | 0.50 | 0.93 | 1.59 | 0.28 | 0.06 | 0.06 |
|  | Median | 26.50 | 23.38 | 521.00 | 3.46 | 3.77 | 15.49 | 43.01 | 0.00 | 0.00 | 2.97 |
|  | IQR | 19.00 | 22.96 | 504.00 | 3.16 | 3.61 | 15.02 | 42.70 | 0.00 | -0.08 | 2.95 |
|  |  | 38.00 | 23.82 | 539.00 | 3.64 | 4.08 | 16.17 | 43.71 | 0.00 | 0.00 | 3.02 |
| Low Myopia (31) | Mean | 26.48 | 24.42 | 533.90 | 3.65 | 3.61 | 16.63 | 44.37 | -3.08 | -0.01 | 3.20 |
|  | SD | 7.48 | 0.74 | 26.94 | 0.29 | 0.27 | 0.80 | 1.20 | 1.75 | 0.02 | 0.13 |
|  | Median | 26.00 | 24.50 | 528.50 | 3.75 | 3.65 | 16.57 | 44.47 | -2.50 | 0.00 | 3.24 |
|  | IQR | 22.00 | 24.14 | 518.00 | 3.50 | 3.44 | 16.22 | 43.36 | -4.50 | 0.00 | 3.11 |
|  |  | 27.00 | 24.70 | 549.00 | 3.82 | 3.72 | 17.03 | 45.03 | -1.50 | 0.00 | 3.28 |
| High Myopia (32) | Mean | 25.66 | 28.02 | 532.41 | 3.74 | 3.62 | 20.13 | 44.43 | -11.27 | 0.06 | 3.69 |
|  | SD | 4.34 | 2.12 | 28.82 | 0.27 | 0.21 | 2.14 | 1.07 | 4.37 | 0.09 | 0.27 |
|  | Median | 26.00 | 27.46 | 530.00 | 3.71 | 3.61 | 19.34 | 44.01 | -10.00 | 0.00 | 3.59 |
|  | IQR | 22.50 | 26.53 | 510.50 | 3.61 | 3.50 | 18.74 | 43.55 | -14.01 | 0.00 | 3.48 |
|  |  | 27.50 | 29.12 | 550.50 | 3.97 | 3.65 | 21.20 | 45.29 | -7.88 | 0.10 | 3.86 |
| Overall (93) | Mean | 28.02 | 25.38 | 530.31 | 3.59 | 3.71 | 17.55 | 43.95 | -4.88 | 0.01 | 3.30 |
|  | SD | 10.79 | 2.43 | 27.65 | 0.35 | 0.37 | 2.42 | 1.44 | 5.55 | 0.07 | 0.35 |
|  | Median | 26.00 | 24.63 | 528.00 | 3.65 | 3.64 | 16.78 | 43.86 | -3.50 | 0.00 | 3.24 |
|  | IQR | 22.00 | 23.76 | 511.00 | 3.42 | 3.49 | 15.81 | 43.10 | -8.00 | 0.00 | 3.02 |
|  |  | 30.00 | 26.87 | 549.00 | 3.78 | 3.74 | 18.76 | 44.95 | 0.00 | 0.00 | 3.50 |
|  | p-Value | 0.64‡ | <0.00‡ | 0.36* | <0.00* | <0.00‡ | <0.00‡ | <0.00* | <0.00‡ | <0.00‡ | <0.00* |

*One-Way ANOVA, ‡Kruskal Wallis, †Dimensionless*

The mean ± standard deviation of all the FAZ dimensions quantified using the new automated method (NAM) are tabulated in Table 2. The result suggests that a narrowing of all the dimensions on comparison between emmetropic vs myopic eyes. Out of 15 dimensions, 10 dimensions showed a significant difference, especially in the case of emmetropic vs high myopic eyes.

*Table 2: Superficial retinal (ILM-IPL) layer's FAZ dimensions (mean ± SD) by NAM*

|  | Normal (30) | Low Moderate myopia (31) | High myopia (32) | p-Value* |
|---|---|---|---|---|
| **Area (mm2)** | 0.31 ± 0.11 | 0.29 ± 0.09 | 0.21 ± 0.08 | <0.00 |
| **Diameter (mm)** | 0.62 ± 0.11 | 0.60 ± 0.10 | 0.51 ± 0.10 | <0.00 |
| **Major axis length (mm)** | 0.69 ± 0.13 | 0.67 ± 0.11 | 0.59 ± 0.15 | 0.01 |
| **Minor axis length (mm)** | 0.58 ± 0.10 | 0.55 ± 0.10 | 0.46 ± 0.08 | <0.00 |
| **Perimeter (mm)** | 2.26 ± 0.46 | 2.08 ± 0.37 | 1.77 ± 0.43 | <0.00 |
| **Eccentricity (mm)** | 0.08 ± 0.02 | 0.07 ± 0.03 | 0.07 ± 0.04 | 0.22 |
| **Fmin (mm)** | 0.59 ± 0.11 | 0.56 ± 0.10 | 0.47 ± 0.09 | <0.00 |
| **Fmax (mm)** | 0.74 ± 0.14 | 0.71 ± 0.12 | 0.63 ± 0.16 | 0.01 |
| **Inner circle radius (mm)** | 0.26 ± 0.05 | 0.25 ± 0.05 | 0.21 ± 0.04 | 0.00 |
| **Circumcircle radius (mm)** | 0.37 ± 0.07 | 0.36 ± 0.06 | 0.32 ± 0.06 | 0.01 |
| **Orientation ($^0$)** | -14.83 ± 43.58 | -5.40 ± 46.19 | -20.60 ± 41.13 | 0.38 |
| **Tortuosity†** | 1.44 ± 0.15 | 1.49 ± 0.14 | 1.45 ± 0.12 | 0.34 |
| **VAD†** | 0.41 ± 0.04 | 0.40 ± 0.05 | 0.40 ± 0.05 | 0.58 |
| **VDI (mm)** | 0.03 ± 0.00 | 0.03 ± 0.00 | 0.03 ± 0.00 | 0.60 |
| **Circularity index†** | 0.77 ± 0.13 | 0.81 ± 0.08 | 0.84 ± 0.10 | 0.03 |

†dimensionless, *One-Way ANOVA

Table 3 shows a comparison of FAZ dimensions quantified by three different methods. On statistical comparison, insignificant differences were observed between the ground-truth (CEM) and the new automated method (NAM). However, a significant difference was noted between the ground-truth (CEM) and inbuilt algorithm method (IAM).

Table 3: A comparison table for FAZ dimension quantified by three different methods.

| | NAM | CEM | IAM | p-Value* |
|---|---|---|---|---|
| **Area (mm2)** | 0.27 ± 0.10 | 0.29 ± 0.10 | 0.24 ± 0.09 | 0.00 |
| **Diameter (mm)** | 0.58 ± 0.11 | 0.60 ± 0.10 | 0.54 ± 0.10 | 0.00 |
| **Major axis length (mm)** | 0.65 ± 0.14 | 0.67 ± 0.11 | 0.62 ± 0.13 | 0.03 |
| **Minor axis length (mm)** | 0.53 ± 0.11 | 0.56 ± 0.11 | 0.50 ± 0.10 | <0.00 |
| **Perimeter (mm)** | 2.03 ± 0.46 | 2.23 ± 0.51 | 1.93 ± 0.44 | <0.00 |
| **Eccentricity (mm)** | 0.07 ± 0.03 | 0.09 ± 0.04 | 0.07 ± 0.03 | 0.01 |
| **Fmin (mm)** | 0.54 ± 0.11 | 0.58 ± 0.11 | 0.51 ± 0.10 | <0.00 |
| **Fmax (mm)** | 0.69 ± 0.15 | 0.73 ± 0.13 | 0.66 ± 0.13 | 0.00 |
| **Inner circle radius (mm)** | 0.24 ± 0.05 | 0.25 ± 0.05 | 0.23 ± 0.05 | <0.00 |
| **Circumcircle radius (mm)** | 0.35 ± 73.79 | 0.37 ± 0.07 | 0.33 ± 0.07 | 0.00 |
| **Orientation ($^0$)** | -13.7 ± 43.64 | 1.81 ± 43.01 | 2.45 ± 42.02 | 0.022 |
| **Tortuosity†** | 1.46 ± 0.13 | 1.43 ± 0.10 | 1.45 ± 0.13 | 0.21 |
| **VAD†** | 0.40 ± 0.05 | 0.41 ± 0.04 | 0.41 ± 0.05 | 0.37 |
| **VDI (mm)** | 0.03 ± 0.00 | 0.03 ± 0.00 | 0.03 ± 0.00 | 0.05 |
| **Circularity index†** | 0.81 ± 0.11 | 0.74 ± 0.13 | 0.80 ± 0.12 | <0.00 |

*One-Way ANOVA*

The Pearson correlation analysis revealed a positive and robust correlation between CEM vs NAM and CEM and IAM (Table 4). However, almost all the FAZ dimensions showed a high correlation coefficient (r) value between CEM and NAM.

Table 4: Correlation between CEM vs NAM and CEM vs IAM

| | NAM (r) | p-Value | IAM (r) | p-Value |
|---|---|---|---|---|
| **Area (mm2)** | 0.793 | 0.000 | 0.560 | <0.00 |
| **Diameter (mm)** | 0.783 | 0.000 | 0.561 | <0.00 |
| **Major axis length (mm)** | 0.669 | 0.000 | 0.327 | 0.00 |
| **Minor axis length (mm)** | 0.808 | 0.000 | 0.638 | <0.00 |
| **Perimeter (mm)** | 0.660 | 0.000 | 0.418 | <0.00 |
| **Eccentricity (mm)** | 0.358 | 0.001 | 0.205 | 0.05 |
| **Fmin (mm)** | 0.668 | 0.000 | 0.606 | <0.00 |
| **Fmax (mm)** | 0.629 | 0.000 | 0.328 | 0.00 |
| **Inner circle radius (mm)** | 0.762 | 0.000 | 0.621 | <0.00 |
| **Circumcircle radius (mm)** | 0.641 | 0.000 | 0.346 | 0.00 |
| **Orientation ($^0$)** | 0.140 | 0.180 | 0.387 | <0.00 |
| **Tortuosity†** | 0.277 | 0.007 | 0.366 | <0.00 |
| **VAD†** | NA | - | NA | - |
| **VDI (mm)** | NA | - | NA | - |
| **Circularity index†** | 0.181 | 0.083 | 0.078 | 0.46 |

*\*\* Correlation is significant at the 0.01 level*

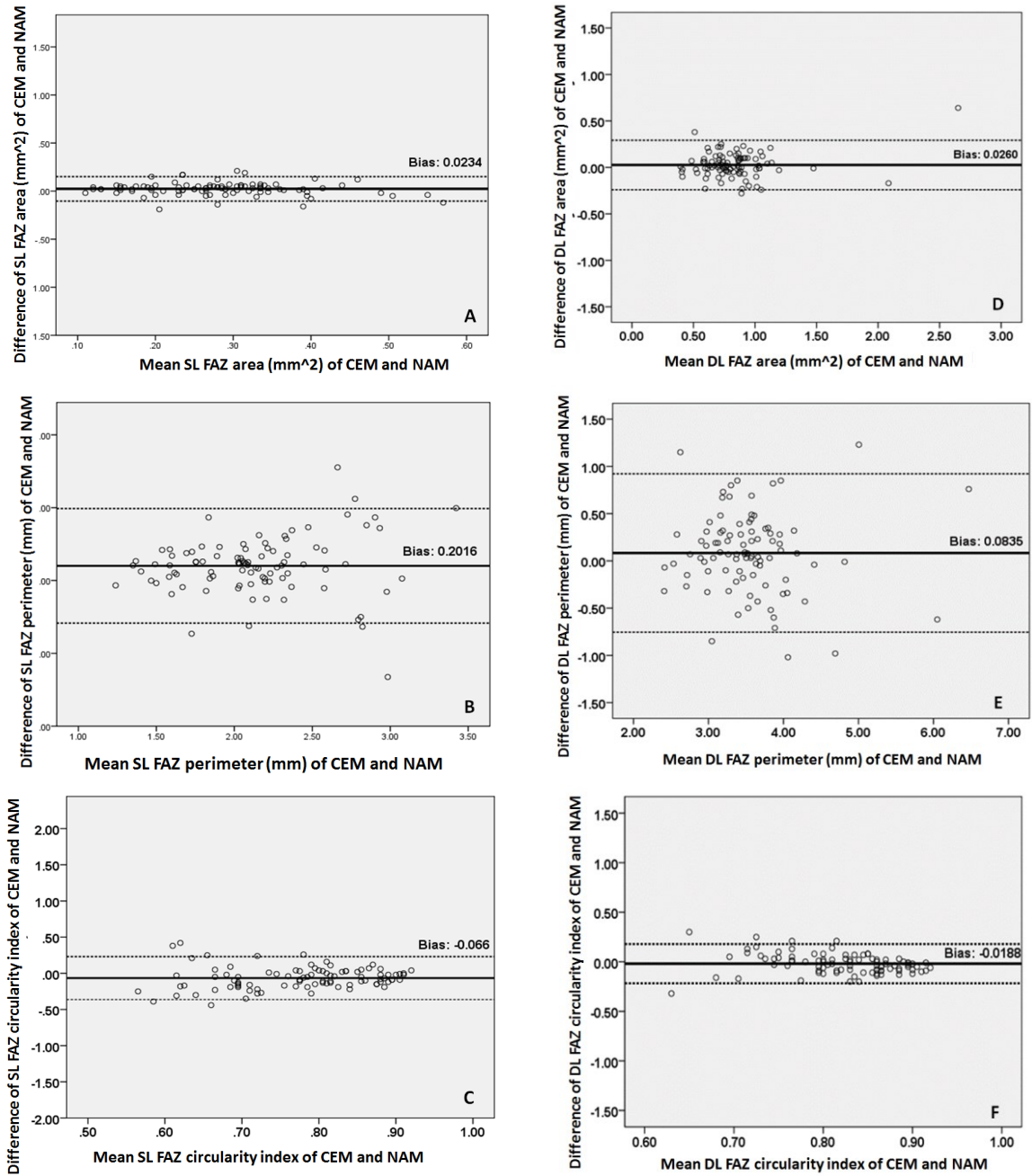

*Fig 4: Bland-Altman plots show an agreement between CEM and NAM for both superficial and deep layers.*

Tables 5 and 6 deal with FAZ dimensions related to the deep layers. Table 5 presents the deep layer's FAZ dimensions quantified by NAM and compared between groups. The results suggest that similar to the superficial layer, the deep layer also narrowed all dimensions when comparing emmetropic vs myopic eyes. There were insignificant differences between various dimensions of emmetropes, low-moderate, and high myopes. Table 6 shows a comparison between CEM and NAM of FAZ dimensions. On statistical comparison, insignificant differences were observed between the ground-truth (CEM) and the new automated method (NAM). However, other dimensions like Eccentricity, $F_{min}$, Tortuosity, VAD showed significant differences. To illustrate the agreement between the two methods (NAM and CEM), Bland-Altman analysis was performed (Fig 4) for both superficial (Fig 4 A, B, & C) and deep layers (Fig 4 D, E, & F). The NAM showed an overall good agreement with CEM. However, the lowest bias was with FAZ area (Fig 4 A & D) and circularity index (Fig 4 C & F). A low to a poor agreement was shown in FAZ perimeter in both superficial and deep layers (Fig 4 B & E).

*Table 5: deep layers FAZ dimension by NAM in different refractive error*

| | Emmetropia (30) | Low-Moderate Myopia (31) | High-Myopia (32) | p-value |
|---|---|---|---|---|
| **Area (mm$^2$)** | 0.93 ± 0.43 | 0.82 ± 0.22 | 0.76 ± 0.23 | 0.16* |
| **Diameter (mm)** | 1.07 ± 0.20 | 1.01 ± 0.13 | 0.97 ± 0.16 | 0.07 |
| **Major Axis Length (mm)** | 1.22 ± 0.26 | 1.16 ± 0.16 | 1.20 ± 0.24 | 0.60 |
| **Minor Axis Length (mm)** | 0.96 ± 0.16 | 0.91 ± 0.14 | 0.82 ± 0.14 | 0.00 |
| **Perimeter (mm)** | 3.73 ± 0.74 | 3.52 ± 0.50 | 3.49 ± 0.72 | 0.29 |
| **Eccentricity (mm)** | 0.14 ± 0.05 | 0.14 ± 0.04 | 0.17 ± 0.06 | 0.09 |
| **$F_{min}$ (mm)** | 0.99 ± 0.18 | 0.93 ± 0.14 | 0.85 ± 0.15 | 0.00 |
| **$F_{max}$ (mm)** | 1.29 ± 0.28 | 1.22 ± 0.17 | 1.24 ± 0.23 | 0.51 |
| **Inner Circle Radius (mm)** | 0.45 ± 0.08 | 0.43 ± 0.06 | 0.38 ± 0.06 | 0.00 |
| **Circum Circle Radius (mm)** | 0.65 ± 0.14 | 0.61 ± 0.08 | 0.62 ± 0.12 | 0.50 |
| **Orientation (Degrees)** | 1.01 ± 27.86 | -4.98 ± 16.23 | -4.27 ± 12.28 | 0.44 |
| **Tortuosity†** | 1.30 ± 0.08 | 1.30 ± 0.08 | 1.28 ± 0.08 | 0.55 |
| **VAD†** | 0.38 ± 0.05 | 0.38 ± 0.03 | 0.32 ± 0.06 | <0.00 |
| **VDI (mm)** | 0.02 ± 0.00 | 0.02 ± 0.00 | 0.02 ± 0.00 | 0.60 |
| **Circularity†** | 0.82 ± 0.06 | 0.82 ± 0.05 | 0.78 ± 0.09 | 0.06 |

†Dimensionless, *The Kruskal-Wallis test

*Table 6: deep layers FAZ dimension comparison between CEM and NAM*

| | CEM | NAM | p-Value |
|---|---|---|---|
| **Area (mm$^2$)** | 0.82 ± 0.29 | 0.85 ± 0.32 | 0.38* |
| **Diameter (mm)** | 1.01 ± 0.16 | 1.02 ± 0.17 | 0.48** |
| **Major Axis Length (mm)** | 1.18 ± 0.23 | 1.19 ± 0.22 | 0.50* |
| **Minor Axis Length (mm)** | 0.88 ± 0.15 | 0.90 ± 0.16 | 0.11* |
| **Perimeter (mm)** | 3.50 ± 0.68 | 3.59 ± 0.66 | 0.44* |
| **Eccentricity (mm)** | 0.13 ± 0.06 | 0.15 ± 0.05 | 0.05* |
| **$F_{min}$ (mm)** | 0.90 ± 0.16 | 0.93 ± 0.17 | 0.05* |
| **$F_{max}$ (mm)** | 1.23 ± 0.24 | 1.25 ± 0.23 | 0.32* |
| **Inner Circle Radius (mm)** | 0.42 ± 0.07 | 0.42 ± 0.07 | 0.56* |
| **Circum Circle Radius (mm)** | 0.62 ± 0.12 | 0.63 ± 0.11 | 0.28* |
| **Orientation (Degrees)** | -0.01 ± 0.01 | -2.53 ± 20.60 | 0.50* |
| **Tortuosity†** | 1.38 ± 0.13 | 1.29 ± 0.08 | 0.00** |
| **VAD†** | 0.38 ± 0.05 | 0.36 ± 0.05 | 0.01* |
| **VDI (mm)** | 0.02 ± 0.00 | 0.02 ± 0.00 | 0.97** |
| **Circularity†** | 0.83 ± 0.09 | 0.81 ± 0.07 | 0.11** |

†Dimensionless, *Mann-Whitney U Test, **Student t-Test

The accuracy of superficial layer FAZ segmentation was done by the clinical expert (JJB) by physically verifying FAZ images segmented by both the inbuilt algorithm method (IAM) and the new method (NAM). Out of 93 images, 58 images (62.37 %) were accurately segmented by the IAM, 10 images (10.75 %) were underestimated, and 25 images (26.88 %) were overestimated. The top two causes for poor segmentation of FAZ were motion artifcats due to instability fixation and poor contrast involving the FAZ. Table 7 displays factors affecting the FAZ segmentation by IAM.

Table 7: FAZ segmentation by IAM and relationship with Signal strength, AXL, SE, and BCVA

| | Underestimation | Optimal | Overestimation | p-Value |
|---|---|---|---|---|
| **Sample size (n)** | 10 / 10.75 % | 58 / 62.37 % | 25 / 26.88 % | |
| **Signal strength** | 10.00 (9.00 – 10.00) | 10.00 (10.00 – 10.00) | 10.00 (9.00 – 10.00) | 0.01 |
| **AXL (mm)** | 26.94 (24.46 – 27.66) | 24.52 (23.79 – 25.72) | 24.75 (23.58 – 27.51) | 0.09 |
| **SE (D)** | -7.38 (0.00 - -13.38) | -2.13 (0.00 - -6.38) | -5.50 (-0.38 - -11.00) | 0.14 |
| **BCVA** | 0.00 (0.00 – 0.20) | 0.00 (0.00 – 0.00) | 0.00 (0.00 – 0.00) | 0.02 |

To evaluate the performance of the segmentations achieved from the NAM algorithm, the Dice similarity coefficient (DSC), a measure of volumetric overlap, was calculated. The Structural Similarity Index (SSIM) was also calculated to determine the similarity between FAZ segmentation by CEM and the other two methods. The DSC and SSIM value which are close to 1.00 indicates high structural similarity. The NAM had an accuracy of 91.40 % (85 eyes). Out of 93 FAZ images, 3 images (3.23 %) were underestimated, and 5 (5.38 %) images were overestimated. Further, The robustness of the calculated FAZ segmentation was determined. These were analyzed using the DSC and the SSIM estimates. The NAM for superficial layer FAZ gave a DSC score of 0.94 and SSIM of 0.97. On the other hand, the NAM for deep layers FAZ gave a DSC score of 0.96 and SSIM of 0.98.

**Discussion**

OCTA is now used for imaging and making dimensional measurements for superficial layer FAZ. However, the inbuilt algorithm only provides restricted parametric parameters, namely area, perimeter, and circularity. On the other hand, the new approach provides the area, perimeter, and circularity and another extra 12 variables that are valuable to clinicians and researchers. Another problem with the built-in algorithm is that it fails to provide consistency of dimensions between commercially available OCTA instruments, i.e. different OCTA instruments do not have similar parametric dimensions used to quantify the FAZ; thus, the parametric dimensions cannot be compared. Furthermore, the inbuilt algorithm does not measure deep layer FAZ, thus making deep layer FAZ analysis difficult for researchers and clinicians with existing instruments. The proposed methodology is available as open source for clinicians and researchers, allowing deep and superficial FAZ en-face images processing. The NAM algorithm was tested for clinically normal and varying degrees of the myopic dataset and found that the results are comparable to the published ones. The new method has certain limitations since it has only been evaluated on average, myopic images and not on other ethnicities or disease conditions.

Additionally, the Bennett axial magnification correction[14,23] was not performed at this stage. There are also ambiguities in circularity and tortuosity, where jitter from manual marking causes differences. In conclusion, these parametric FAZ dimensions can be considered biomarkers for myopic conditions using the FAZ dimensions. In the future, this work can be extended to other conditions. The existence of capillaries in and around the FAZ in the superficial layer causes abruptness, which causes the algorithm to fail in some cases for correctly detecting the FAZ in the superficial layer FAZ. As a result, the option of using higher resolution FAZ images to improve their accuracy can be explored.

**Conclusion**

A clinician-friendly, publicly available, open-access algorithm with an easy to use GUI was constructed and tested on the deep and superficial FAZ images. The new algorithm outperforms the inbuilt algorithm both in the case of superficial layer and deep layers images. This open GUI software is online at the https://github.com/VIEKASH2001/FAZSeg and can validate newly developed automated segmentation algorithms by others.

**List of abbreviations**

ACD: Anterior chamber depth; ACR: axial length/corneal radius ratio; AL: axial length; BCVA: The best-corrected visual acuity; CCT: Central corneal thickness; CEM: Clinical expert method; D: dioptre; DR: Diabetic Retinopathy; FAZ: The Foveal Avascular Zone; GAT: Goldmann applanation tonometer; GUI: Graphical User Interface; IAM: Inbuilt algorithm method; LT: Lens thickness; mm: Millimetre; MSE: Mean spherical equivalent; NAM: New automated method; OCT: Optical coherence tomography; OCTA: Optical coherence tomography angiography; SSIM: Structural Similarity Index. VAD: Vessel avascular density; VDI: Vessel diameter index; WTW: White to white;

**Declarations**

The authors report no conflicts of interest in this work.

**Ethics approval and consent to participate**

The Institutional Review Board approved this study of the Vision Research Foundation, Chennai, India. The study conformed to the tenets of the Declaration of Helsinki, and signed informed consent was obtained from all subjects.

**Consent for publication**

Not applicable

**Availability of data and materials**

All the clinical data and materials supporting the manuscript are maintained in our Hospital.

**Competing interests**

The authors declare that they have no competing interests.

**Funding**

None

**Acknowledgements**

This work was partly supported by a DISCOVERY Grant from the Natural Sciences and Engineering Research Council of Canada to V. L.

# Appendix

*Table A1: GUI Buttons, tunable parameters and their functions*

| GUI Button | Tuneable Parameter | Operation | Description |
|---|---|---|---|
| Select Directory | - | Choosing dataset directory | Selects the directory containing the dataset images |
| Manual | - | Manual segmentation | Optional operation used to manually segment the FAZ |
| Load | Crop X | Load image, complement and crop | Crops the area extended by the top and right image portions |
| | Crop Y | | Crops the area extended by bottom and left image portions |
| Threshold | Fudge | Prewitt's thresholding | Thresholding value for the cropped region |
| Close | Dilation | Image Closure | Length of linear structuring element that is used for dilation |
| | Erosion | | Radius of disk-shaped structuring element used for erosion |
| Segment | Filled | Segmentation | Visualizing the segment via region infilling |
| | Outline | | Visualizing the segment via region outline |
| Generate | - | Generate measurements | Perform measurements of 15 parametric dimensions |

| Export | - | Export measurements | Export the generated measurements to a .csv file |